\documentclass[%
 reprint,
 onecolumn,
 amsmath,amssymb,
 aps,
]{revtex4-2}
\usepackage{graphicx}
\usepackage{bm}
\usepackage{hyperref}
\usepackage{color}

\begin{document}


\title{Splitting the Nariai cosmology}

\author{Mustafa Halilsoy}
 \altaffiliation[]{mustafa.halilsoy@emu.edu.tr}
\author{Chia-Li Hsieh\footnote{Author to whom any correspondence should be addressed.}}
 \email{galise@gmail.com} 
\affiliation{%
 Department of Physics, Faculty of Arts and Sciences, Eastern Mediterranean University, Famagusta,
North Cyprus via Mersin 10, Turkey 
}%


\date{\today}

\begin{abstract}
At certain time the Nariai spacetime is split into two parts. The resulting cosmology consists of the original Nariai and a new component with topology $(flat)_2\times S^2$, that is equivalent to a cloud of strings. We explore the properties of our hybrid cosmological model.
\end{abstract}

\maketitle



\section{\label{sec:level1}Introduction}

Once a cosmological model such as Kasner \cite{Kasner1921}, de Sitter \cite{de Sitter1917}, Nariai \cite{Nariai1950}, etc. takes a start, does it necessarily preserve the same identity or by some mechanism it transforms into a new identity? The cause of the change may be due to a phase change or coupling with another system, as we have in the system of coupling oscillators. When driven by an external force, the system is open for such changes. Herein our aim is not to consider coupling of huge number of systems such as neurons or oscillators as in the case of a chimera system \cite{Abrams2004}, but concentrate on just a few ones such as cosmologies. The question raised above constitutes our main aim in connection with the well-known cosmological model of Nariai. But similar models can also be developed for any cosmological model of physical interest. We recall that Nariai spacetime is a singularity free, homogeneous, non-isotropic one powered by a cosmological constant. We reiterate that the shear factor, which is absent in de Sitter and Kasner cosmologies entails the Nariai metric and its derivatives regular.Unlike the de Sitter cosmology, it is not conformally flat (CF), that is, it has non-zero Weyl curvature. Since it is sourced by a cosmological constant, all non-zero physical quantities are expressed in terms of that constant. We revise the metric function in the Nariai spacetime such that the metric and its first derivative remains continuous whereas the second derivative is not continuous. This automatically invites new forces, new energy-momentum components and a new spacetime in connection with the original Nariai spacetime. By this method, we show that the Nariai cosmology, beyond certain fixed time (say $t_0$) survives by giving birth to a 'baby cosmos'. The original Nariai spacetime has the topology of $(non-flat)_2\times S^2$ whereas its 'baby cosmos' has the form of $(flat)_2\times S^2$. Further, considering that the time extends from $-\infty<t<+\infty$, and with the symmetry arguments we confine, the original Nariai cosmology to $-t_0<t<+t_0$, leaving the rest $t>|t_0|$ with the 'baby universe'. We add also that our 'baby universe' is completely classical, it has nothing to do with the one discussed in quantum field theory \cite{Strominger1991}. It turns out that what we call the 'baby universe' is identified as the cosmology sourced by a cloud of strings \cite{Letelier1979}. {It should also be supplemented that the idea of string cloud component was considered previously in a different context consisting of a scalar field hedgehog model\cite{Guendelman1991}, and it can be applied to various cosmologies as well.}

Organization of the paper is as follows. In section II, we review the Nariai cosmology in brief. In section III, we apply our technique of revising the underlying spacetime. We complete the paper with our conclusion in section IV.




\section{Nariai spacetime revisited}

Nariai spacetime is expressed by
\begin{equation}
 ds^2=dt^2-\cosh^2(kt)dz^2-\frac{1}{k^2}(d\theta^2+\sin^2\theta d\phi^2),
\end{equation}
where $k$=constant, that can be identified with the cosmological constant $\Lambda=k^2$. The Einstein equation
\begin{equation}
 G_\mu^\nu=R_\mu^\nu-\frac{R}{2}\delta_\mu^\nu
\end{equation}

\begin{equation}
 =-T_\mu^\nu
\end{equation}
are satisfied with the energy-momentum tensor
\begin{equation}
  T_\mu^\nu=k^2 diag(-1, -1, -1, -1).
\end{equation}
This implies from the identification
\begin{equation}
  T_\mu^\nu=diag(-\rho, p_z, p_\theta, p_\phi),
\end{equation}
that we have the energy density $\rho=k^2$ and the pressures $p_z=p_\theta=p_\phi=-k^2$. The weak energy condition (WEC). i.e. $\rho>0, \rho+p_i\geqslant0$ are satisfied. The strong energy condition (SEC), i.e. $\rho+\sum_{i=1}^{3} p_i\geqslant0$ is obviously violated. As a matter of fact, the satisfaction of WEC, or the null-energy condition (NEC) is enough to characterize our system as physical.

The non-affinely parametrized choice of the Newman-Penrose (NP) null-tetrad \cite{Newman1962} is
\begin{equation}
  \begin{array}{c}
  l_\mu=\frac{1}{\sqrt{2}}(\delta_{\mu}^t-\cosh(kt)\delta_{\mu}^z), \\
  n_\mu=\frac{1}{\sqrt{2}}(\delta_{\mu}^t+\cosh(kt)\delta_{\mu}^z), \\
  m_\mu=\frac{1}{\sqrt{2}k}(\delta_{\mu}^\theta+i\sin\theta\delta_{\mu}^\phi), \\
  \bar{m}_\mu=\frac{1}{\sqrt{2}k}(\delta_{\mu}^\theta-i\sin\theta\delta_{\mu}^\phi).
  \end{array}
\end{equation}
This gives the spin coefficients
\begin{equation}
 \begin{array}{c}
   \epsilon =-\gamma=\frac{k}{2\sqrt{2}}\tanh(kt), \\
   \alpha =-\beta=\frac{k}{2\sqrt{2}}\cot\theta.
 \end{array}
\end{equation}
and the only non-zero NP quantities are
\begin{equation}
  \Psi_2=-\frac{k^2}{3},
\end{equation}
with the scalar curvature
\begin{equation}
  R=4k^2,
\end{equation}
The non-zero $\Psi_2$ justifies that the Nariai spacetime is not CF, with the positive scalar curvature and all vanishing Ricci components implying the absence of any source other than the cosmological constant. We add that the absence of the shear in the spacetime is due to the choice of our tetrad. An alternative choice of null-tetrad gives different quantities. For instance, the choice
\begin{equation}
  \begin{array}{c}
  l_\mu=\frac{1}{\sqrt{2}}(\delta_{\mu}^t-\frac{1}{k}\delta_{\mu}^\theta),\\
  n_\mu=\frac{1}{\sqrt{2}}(\delta_{\mu}^t+\frac{1}{k}\delta_{\mu}^\theta),\\
  m_\mu=\frac{1}{\sqrt{2}k}(\cosh(kt)\delta_{\mu}^z+\frac{i}{k}\sin\theta\delta_{\mu}^\phi),\\
  \bar{m}_\mu=\frac{1}{\sqrt{2}k}(\cosh(kt)\delta_{\mu}^z-\frac{i}{k}\sin\theta\delta_{\mu}^\phi).
  \end{array}
\end{equation}
gives the spin coefficients
\begin{equation}
 \begin{array}{c}
   \mu =-\sigma=\frac{k}{2\sqrt{2}}(\tanh(kt)-\cot(\theta)), \\
   \lambda =-\rho=\frac{k}{2\sqrt{2}}(\tanh(kt)+\cot(\theta)).
 \end{array}
\end{equation}
where $\sigma$ stands for the shear and $\rho$ the expansion. For the present choice, we obtain
\begin{equation}
 \begin{array}{c}
   \Psi_2 =\frac{k^2}{6}, \\
   \Psi_0=\Psi_4 =-\frac{k^2}{2},
 \end{array}
\end{equation}
and the scalar curvature $R=4k^2$, unchanged. It is seen that the condition for type-D
\begin{equation}
  9\Psi_2^2=\Psi_0\Psi_4
\end{equation}
is checked. The interest in the shearing and expanding tetrad choice (10) is due to the fact that these quantities arise in the Raychaudhuri equation, which is crucial for the singularity analysis \cite{Raychaudhuri1955}. However, in the present analysis our spacetime is already regular so that we shall not follow the route of that equation.

\section{hybridization of the Nariai cosmology}
Instead of the $\cosh(kt)$ term in the $g_{zz}$ metric function in (1), we impose now
\begin{equation}
  f(t)=\cosh k[t-(t-t_0)\theta(t-t_0)]+k\sinh(kt_0)(t-t_0)\theta(t-t_0),
\end{equation}
where $t_0$ is a constant time and $\theta(t-t_0)$ is the unit step function with standard definition
\begin{equation}
\theta(x) =
\left\{
    \begin{array}{lr}
        1, \qquad   \text{if } &  x > 0\\
        0, \qquad   \text{if } &  x \le 0
    \end{array}
    \right\}
\end{equation}
The function $f(t)$ is chosen in particular to satisfy
\begin{equation}
    f(t) =
    \left\{
    \begin{array}{lr}
        a+bt, \qquad   \text{if } &  t > t_0\\
        \cosh kt, \qquad   \text{if } &  t \le t_0
    \end{array}
    \right\}
\end{equation}
in which the constants $a$ and $b$ are given by
\begin{equation}
 \begin{array}{c}
   a =\cosh kt_0-kt_0\sinh kt_0 \\
   b=k\sinh kt_0.
 \end{array}
\end{equation}

Now, we make the choice $a=0$ and $b=1$, to provide continuity in the metric at $t=t_0$. This admits the possible numerical solutions
\begin{equation}
 \begin{array}{c}
   k \simeq 0.66 \\
   t_0\simeq 1.8
 \end{array}
\end{equation}
and $kt_0\simeq 1.19$. Under the imposition of the metric function (14) obviously we have the standard Nariai cosmology for $t \le t_0$. and a new 'baby universe' described by (for $t>t_0$)
\begin{equation}
  ds^2=dt^2-t^2dz^2-\frac{1}{k^2}(d\theta^2+\sin^2\theta d\phi^2).
\end{equation}
Overall, for $-\infty<t<+\infty$, we have
\begin{equation}
  ds^2=dt^2-f(t)^2dz^2-\frac{1}{k^2}(d\theta^2+\sin^2\theta d\phi^2).
\end{equation}
where we plot $f(t)$ in Fig. 1.
\begin{figure}
  \centering
  \includegraphics[width=0.5\textwidth]{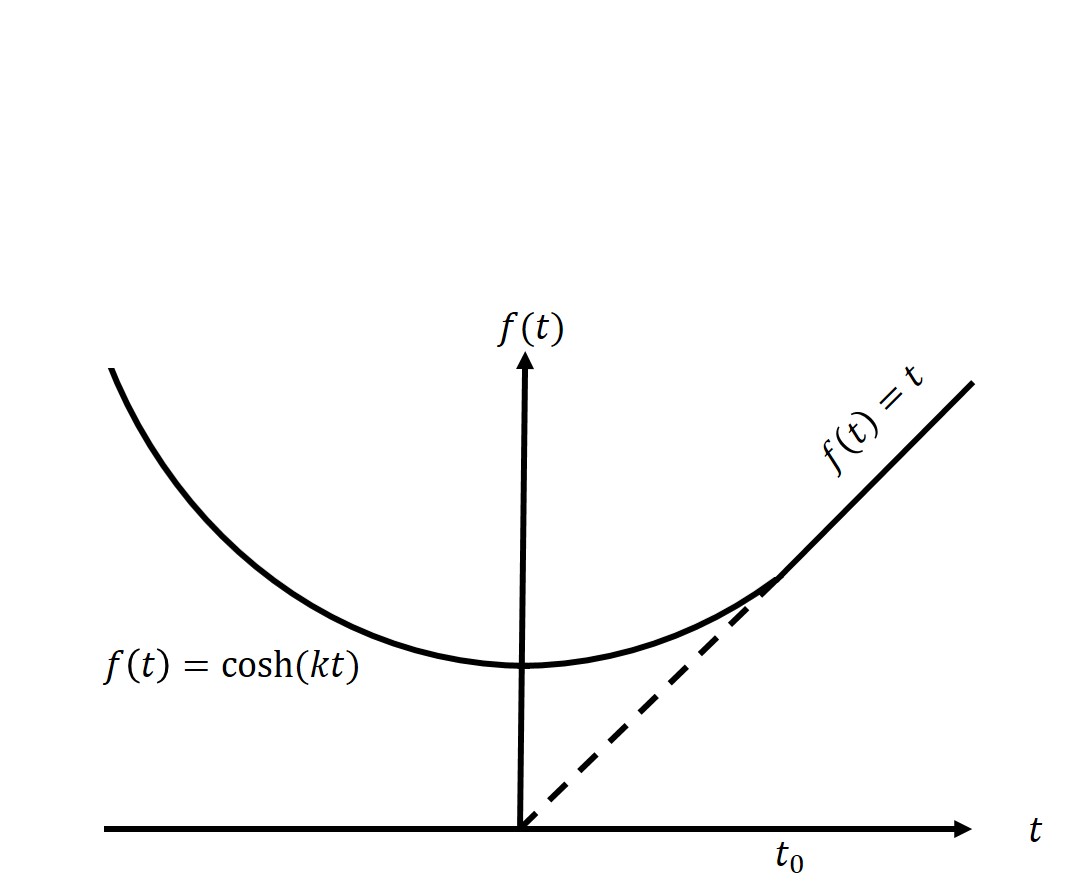}
  \caption{A plot of the function $f(t)$ for $0\le |t|<\infty.$}\label{f1}
\end{figure}

With the new metric, we make the choice of the tetrad as
\begin{equation}
  \begin{array}{c}
  l_\mu=\frac{1}{\sqrt{2}}(\delta_{\mu}^t-t\delta_{\mu}^z), \\
  n_\mu=\frac{1}{\sqrt{2}}(\delta_{\mu}^t+t\delta_{\mu}^z), \\
  m_\mu=\frac{1}{\sqrt{2}k}(\delta_{\mu}^\theta+i\sin\theta\delta_{\mu}^\phi), \\
  \bar{m}_\mu=\frac{1}{\sqrt{2}k}(\delta_{\mu}^\theta-i\sin\theta\delta_{\mu}^\phi).
  \end{array}
\end{equation}
and obtain the non-zero NP scalars as follows
\begin{equation}
  \begin{array}{c}
    \Psi_2=-k^2/6, \\
    \phi_{11}=k^2/4, \\
    R=2k^2.
  \end{array}
\end{equation}
That is, with the choice of the metric function $f(t)$, for $t>0$, the conformal curvature and scalar curvature modify slightly, but more importantly there is an emergent non-zero Ricci tensor. The resulting energy-momentum tensor takes the form
\begin{equation}
  T_\mu^\nu=k^2 diag(-1, -1, 0, 0),
\end{equation}
which suggests that for $t>0$, the 'baby universe' is no more in cosmological form but it represents a cloud of strings \cite{Letelier1979}. Instead, the angular pressures vanish and as expressed in (22), a non-zero Ricci component arises. Still, in the resulting 'baby universe' for $t>0$, the WEC is satisfied. The regularity of (20) is manifest although it is expressed in a non-shearing basis (21). The choice of a basis with shear, which is always possible, will yield more non-zero components of Ricci and curvature.

Finally, we can symmetrize our hybrid cosmology by including the negative time $-\infty<t<0$. This amounts to choosing $t\rightarrow |t|$, so that the resulting function for $f(t)$ modifies as in Fig. 2.
\\

\begin{figure}
  \centering
  \includegraphics[width=0.5\textwidth]{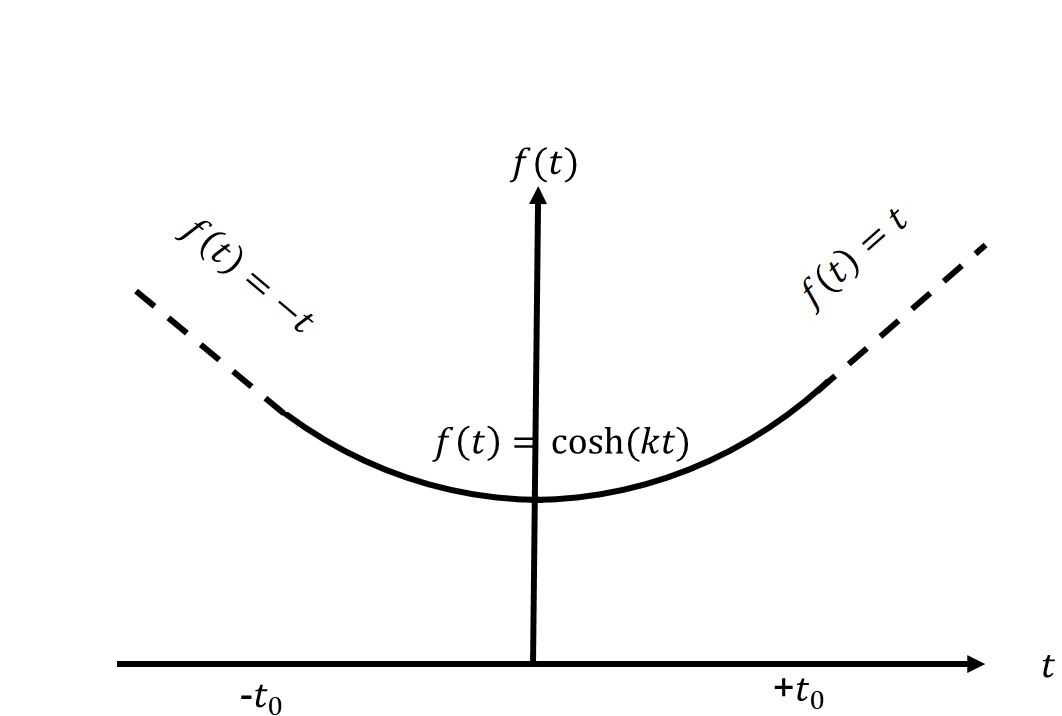}
  \caption{The metric function $f(t)=\cosh kt$ for $|t|<t_0$, and $f(t)=t$ for $|t|>t_0$.}\label{f2}
\end{figure}
\section{Conclusion}
The possibility of reformulating the Nariai cosmology as split into two/three parts has been considered. This is to be done at a classical level so that unlike the ‘rabbit-from hat’ mechanism of quantum, we rely on the geometrical conditions such as continuity of metric and its first derivatives. The second derivatives, however, involves discontinuity since it amounts to exchange energy with outside. The continuity of second derivative should be imposed provided we match a sourceful spacetime with vacuum. The continuity of the extrinsic curvature (i.e. the second fundamental form) becomes necessary in order to avoid source accumulation at the boundary. This was the case in joining the Levi-Civita metric with vacuum \citep{Levi1917, Halilsoy2022}. Since we have a sourceful spacetime to be matched with another sourceful one (of cosmological constant), the redistribution of sources between regions is not unexpected. As a result, the structure of Nariai metric \cite{Dadhich2001} modifies beyond the time $t=\pm t_0$, into a topology of the form $(flat)_2\times S^2$. That is, the cosmological form of Nariai becomes restricted by the time $t<|t_0|$. We interpreted this case as the Nariai cosmology giving birth to a cloud of string universe, which is already a well-accepted model cosmology. Finally, we remind that at a phase transition point we have also the case of a divergent specific heat function. Discontinuity in the stress-energy therefore can be attributed to a spontaneous phase transition in the Nariai cosmology. We remark finally that the method of splitting Nariai which we have employed here is applicable with minor variations to different cosmological models. {Ref. \cite{Guendelman1991} provides one such application to a multi-scalar hedgehog model in a spherically symmetric spacetime.}

Data Availability Statement: No Data associated in the manuscript




\end{document}